\documentclass{appolb}
\usepackage{graphicx}
\usepackage[colorlinks=true,citecolor=blue, ]{hyperref}%

\begin{document}
\title{Evaluation of Energy Resolution by Changing Angle and Position of Incident  Photon  in a LYSO Calorimeter  }
\author{A. M. E. SAAD\thanks{ahmedelfatih43@hotmail.com}, F. KOCAK\thanks{fkocak@uludag.edu.tr}
\address{Department of Physics, Bursa Uludag University, 16059, Bursa, Turkey}}

\maketitle
\begin{abstract}

In this paper, we investigate the effect on energy resolution from changing the angle and the position of incidence photon for a 5 $\times$ 5 crystal matrix built with (25 $\times$ 25 $\times$ 200) mm$^{3}$ LYSO scintillating crystals. Those crystals have been proposed for the electromagnetic calorimeter of the Turkish Accelerator Center-Particle Factory (TAC-PF) detector. The energy resolution was obtained as $\sigma_{E}/E =  0.42 \% / \sqrt{E/GeV} \oplus 1.60 \%$ at the center of the matrix in the energy range of 50 MeV to 2 GeV. When we examined the dependence of the energy resolution on the incidence angle of the photon, resolution began to deteriorate at angles greater than $2^{\circ}$ on the 5 $\times$ 5 crystal matrix. Moreover, energy resolution at the corners of the central crystal was worse than at the center of the central crystal by a factor of 1.3 at 50 MeV and 1.1 at 2 GeV.

\end{abstract}
\section{Introduction}
In the study of elementary particles in high-energy physics, it is required not only to determine which particles are born but also to measure their characteristics with high accuracy, especially their trajectory, momentum, and energy. All this is done using detectors. Tracker detectors measure the trajectory and momentum of particles without introducing any distortion, and the calorimeters completely absorb a particle and measure its energy. In experiments, electromagnetic calorimeters are used to measure total energy and identify particles (including neutral ones). The detection of both charged and neutral particles in a segmented calorimeter allows one to obtain information about the coordinates of particles and showers; to identify particles, for example, by separating photons, electrons, protons, pions, etc.; and to trigger event selection systems. This is because calorimeters can offer fast, easy-to-process and interpretable signals \cite{fabjan2003calorimetry}.\\
High-energy electrons and photons, passing through the substance of the scintillator, collide mainly with electron shells of atoms. They generate an electromagnetic shower of electrons, positrons, and photons. The number of particles in the shower increases rapidly until the average particle energy drops to the lowest level. Research and development of new high-density scintillators began with the Compact Muon Solenoid (CMS) Collaboration after the first announcement of lead tungstate crystal (PWO) \cite{lecoq1999scintillator}. Recently, it was found that the modules of the end parts of the electromagnetic calorimeter based on PWO  crystals were damaged by high-energy hadrons during the operation of the Large Hadron Collider (LHC) \cite{dissertori2010study}. For this reason, consideration has been given to replacing the PWO crystal with a new generation of radiation-resistant crystals. One of the promising scintillators comprises crystals based on lutetium orthosilicate. Lutetium-yttrium oxyorthosilicate (Lu$_{1.8}$Y$_{.2}$SiO$_{5}$:Ce) scintillation crystals (LYSO)  were proposed for use in high-energy physics experiments as promising materials for homogeneous electromagnetic calorimeters \cite{chen2007gamma}. This is because of their high light output (32 000 ph/MeV), short decay time (41 ns), high density (7.10 g/cm$^3$)  , great time and energy resolution, and stable physical and chemical properties. Some of these experiments are the Muon-to-Electron (Mu2e) Experiment at Fermilab \cite{pezzullo2014lyso} and the SuperB experiment in Europe \cite{eigen2013lyso}. It has also been proposed that a LYSO/W/Quartz capillary sampling calorimeter be constructed for the CMS upgrade \cite{zhang2013lso}. The scintillator was also considered for the ECAL part of the proposed TAC-PF detector, in addition to PWO and CsI(Tl) crystals \cite{kocak2017simulation}. In addition, the scintillator has been used in COherent Muon to Electron Transition (COMET) experiment to build a total absorption calorimeter \cite{oishi2014lyso} and for the High Energy cosmic-Radiation Detection (HERD) experiment in space \cite{zhang2014high, hu2019neutron}. One of the main reasons for not considering LYSO crystals in some high-energy physics experiments is the high cost related to its high melting point and the costs of raw materials.
\section{GEANT4 Simulation}



\begin{figure}[htb]
\centerline{%
\includegraphics[width=12.5cm]{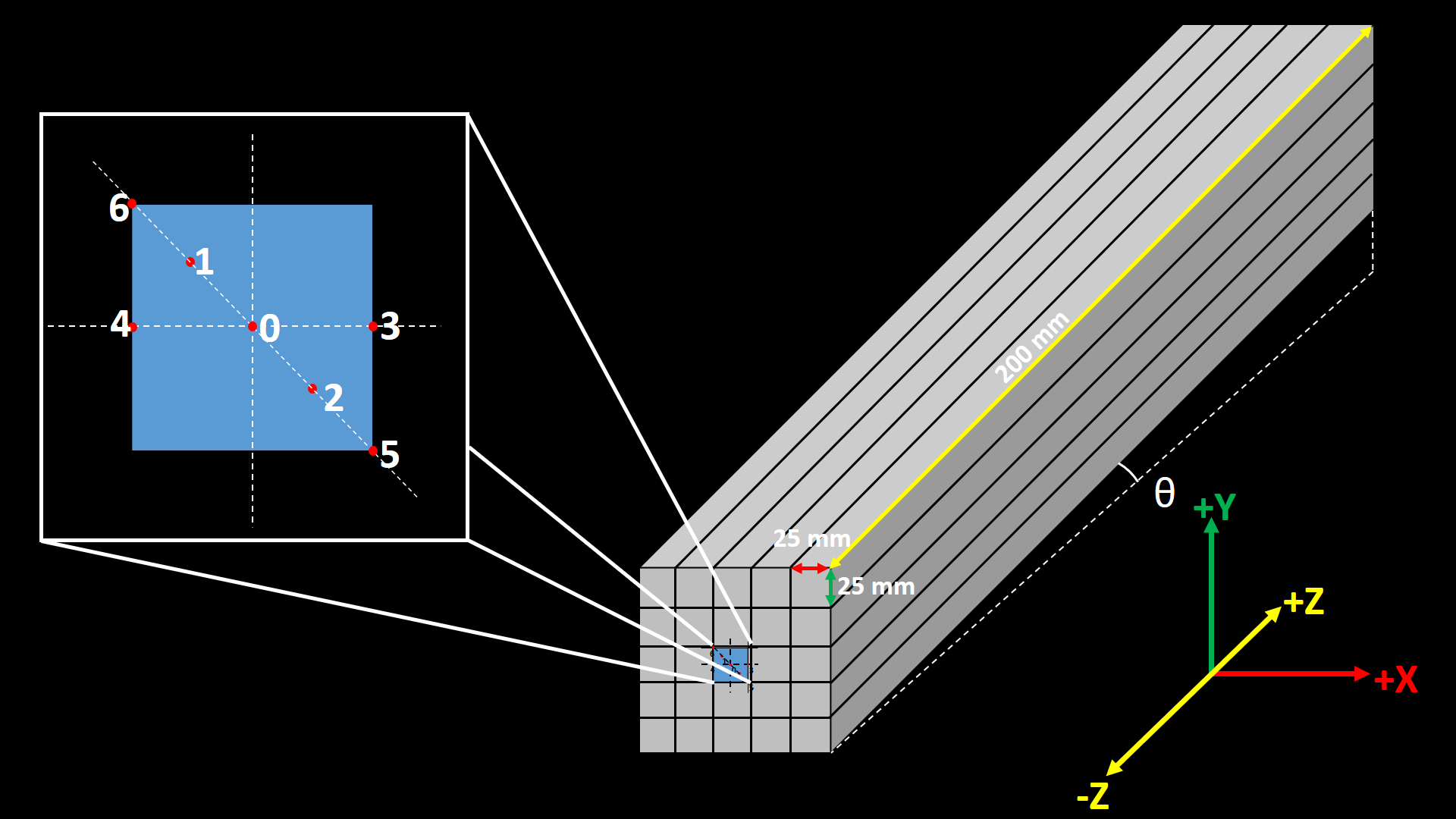}}
\caption{Schematic view of the crystal array of a 5 $\times$ 5  arrangement of LYSO crystals, showing the interaction positions of the incident photon within the central crystal, as well as angle $ \Theta$. The coordinate system used to specify the entry point and the angle of incidence also is  indicated.  }
\label{Fig:crystal}
\end{figure}

GEANT4 is a software package designed for modelling the passage of particles through matter based on the Monte Carlo method \cite{agostinelli2003geant4}. This toolkit is a set of libraries implemented in C++. Configuration of particle sorts included in the simulation, physical processes, models of particle interaction, and their application boundaries are described in a particular class of the program based on GEANT4 PhysicsList. Usually, PhysicsList includes a set of electromagnetic and hadron interactions, the decay of nuclei and particles, and parameterized interaction models. However, GEANT4 allows very flexible use of physical interaction model particles with matter. A researcher can choose certain processes and interaction models independent of simulation requirements. For modeling particle showers, GEANT4 provides various lists (sets) of physical processes. One of them, $emstandard$\_$opt1$, was selected for this study. $emstandard$\_$opt1$ is used to describe high-energy interactions \cite{ribon2010status}. For a better understanding of the calorimeter and its ability to predict behavior at various angles and levels of energy, the simulation was performed using the GEANT4.10.4.3 on a 5 $\times$ 5 LYSO crystal matrix with a length of 200 mm (17.5 X$_{0}$) and 25 $\times$ 25 mm$^2$ (1.2 R$_{\mathcal{M}}$) cross section. All the simulation results were based on the energy deposition in the scintillator. A photon at eight different values of energy from 50 MeV to 2.0 GeV was injected at different angles to the central crystal along the Z-axis from 0.01 mm away (see Fig. \ref{Fig:crystal} for the definition of the coordinates).

\begin{figure}[htb]
\centerline{%
\includegraphics[width=12.5cm]{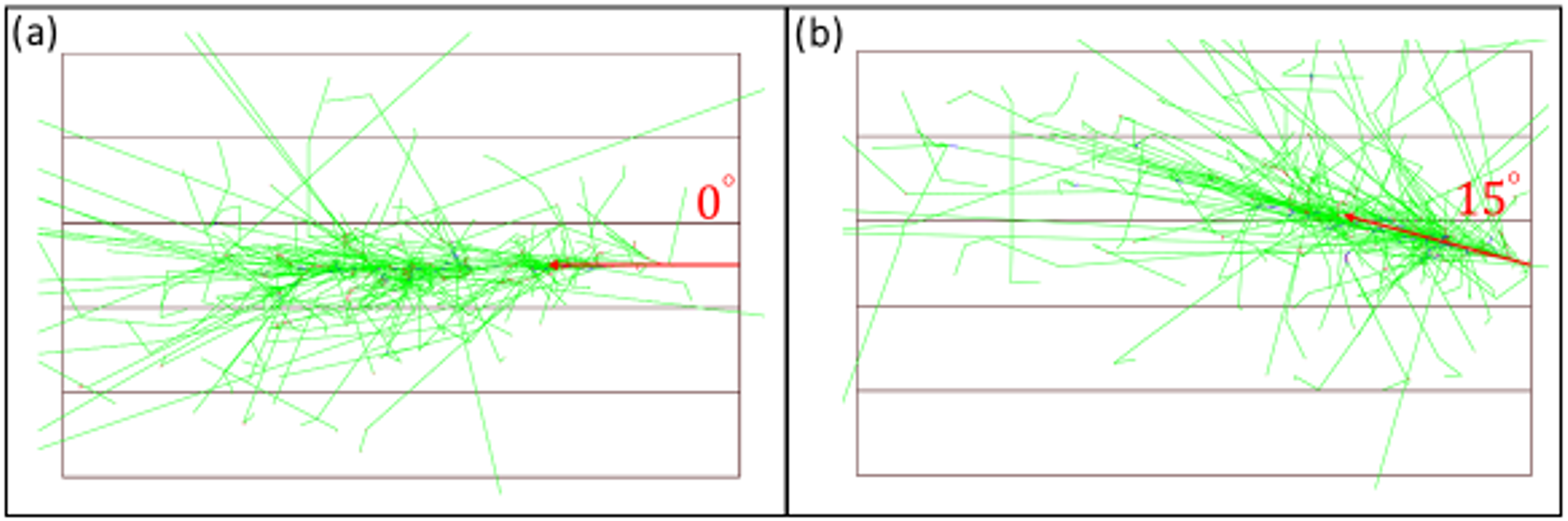}}
\caption{Development of an electromagnetic cascade of a 5 $\times$ 5 LYSO matrix for a single incident photon at 1 GeV with (a)   $ \Theta$ = $0^{\circ}$ and (b)  $ \Theta$ = $15^{\circ}$ along the Z-axis. }
\label{Fig:shower}
\end{figure}

Moreover, a simulation was performed to study the dependence of the energy resolution  on the different interaction positions of the incident particle, as shown in  Fig. \ref{Fig:crystal}. For example, Fig. \ref{Fig:shower} shows the development of an electromagnetic shower in a 5 $\times$ 5 LYSO crystal matrix  at 1 GeV for two different photon incident angles. Most of its energy is deposited in the crystal matrix. The rest of the energy is electromagnetic shower leaks. 

\section{Result and Discussion}
\begin{figure}[htb]
\centerline{%
\includegraphics[width=12.5cm]{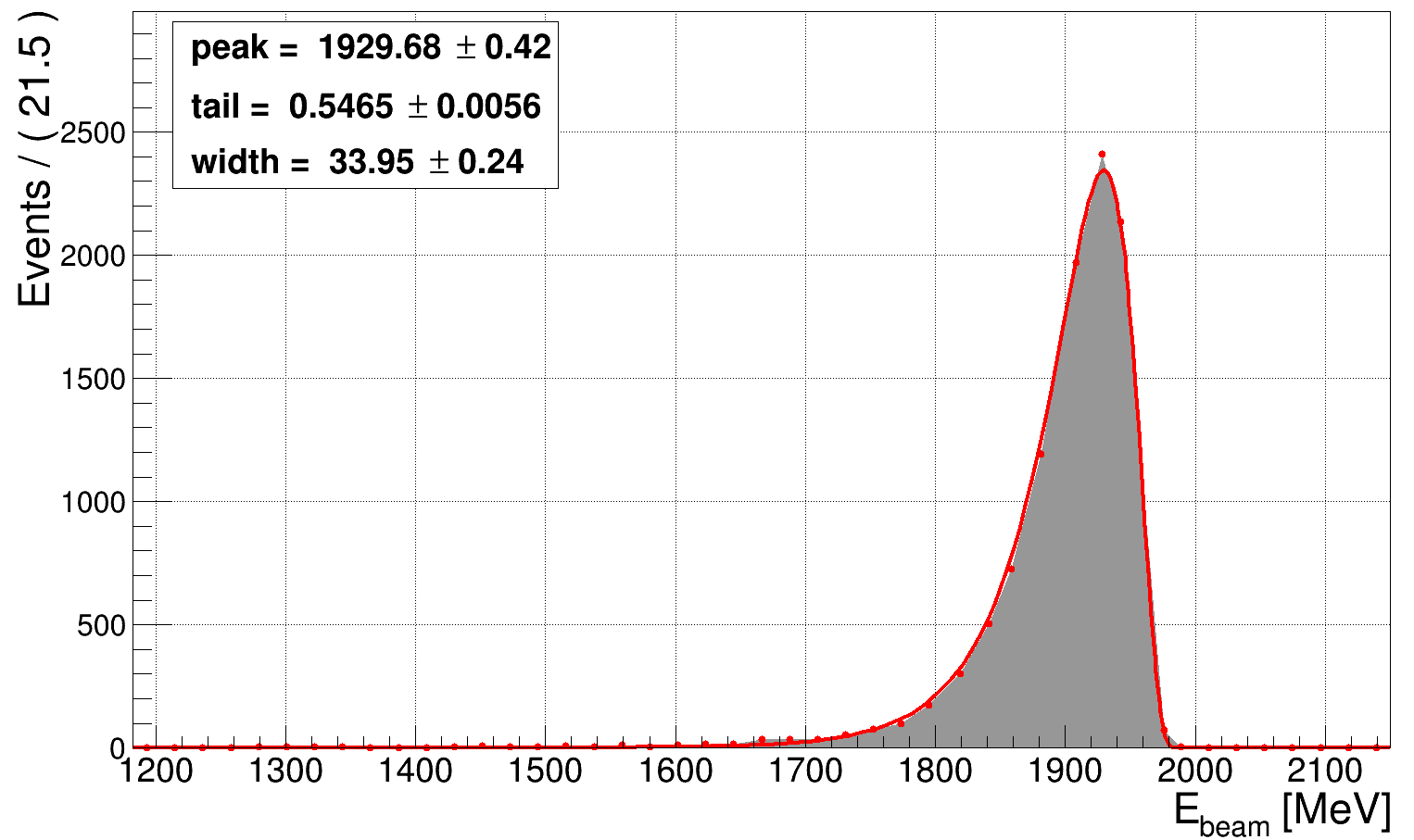}}
\caption{Response of the LYSO calorimeter for 2.0 GeV photon. The red curve indicates a fit with a Novosibirsk function.}
\label{Fig:spectra}
\end{figure}

\begin{figure}[htb]
\centerline{%
\includegraphics[width=12.5cm]{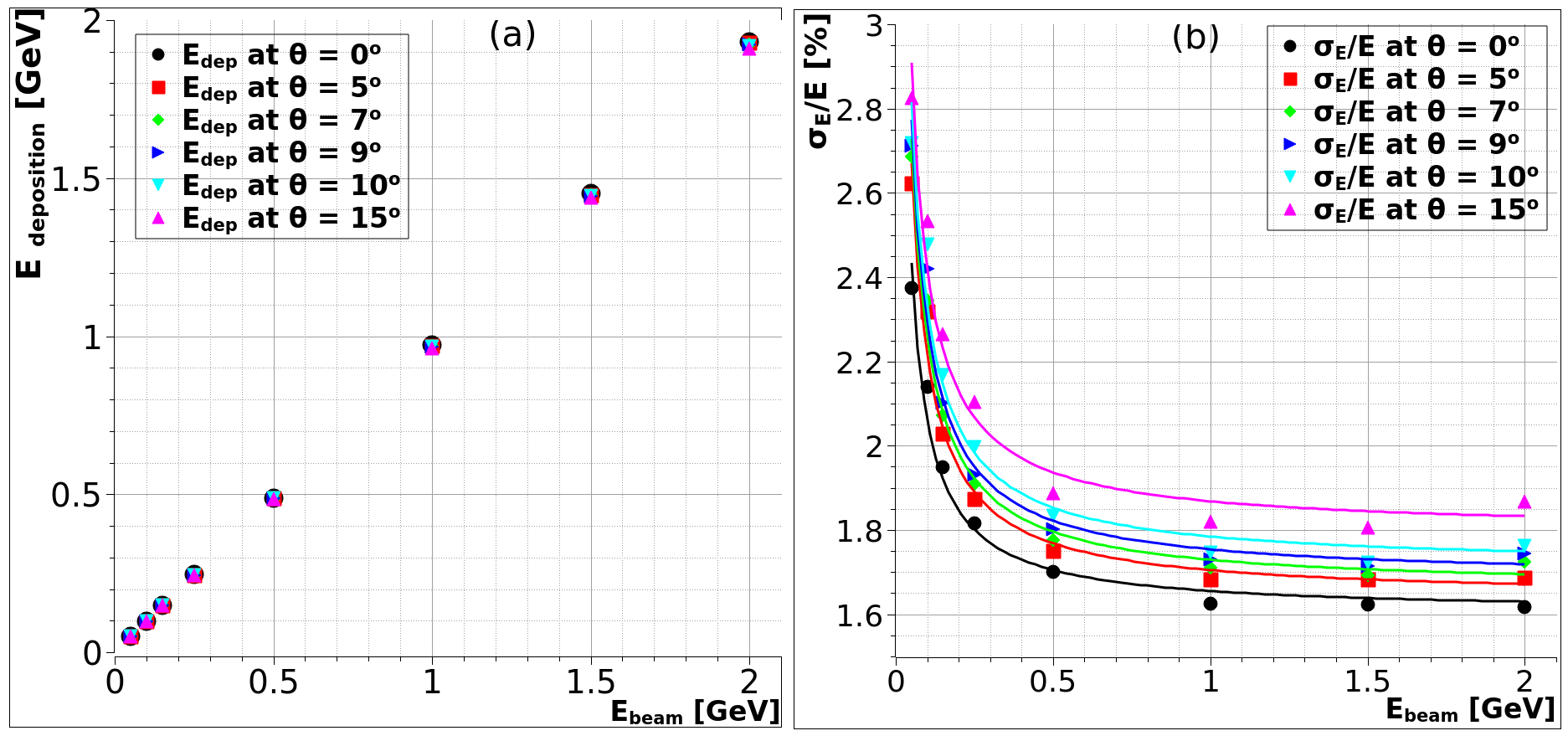}}
\caption{Dependence of the energy deposit (a) and energy resolution (b) on the energy of photons entering the calorimeter at different angles.   }
\label{Fig:2in1}
\end{figure}

The electromagnetic calorimeter was designed to measure the energy and position of high-energy gamma rays, as well as electrons and positrons. Photons in the energy range from 50 MeV to 2 GeV were injected into the center of central crystal of the matrix. The energy released in each  crystal was recorded. The energy range was determined according to the requirement of the TAC-PF ECAL. To determine the effect of the angle of incidence of the particles on energy resolution, the photon was injected into the central crystal of the matrix at different angles between $0^{\circ}$ and $15^{\circ}$. The total energy spectra for the incident photons  were obtained by summing  all the energies deposited in each crystal. Because the Gaussian form of energy deposition spectra has  an asymmetric tail towards lower energies, the distribution was fitted to a Novosibirsk function to determine energy resolution  \cite{ikeda2000detailed}. For example, Fig. \ref{Fig:spectra} illustrates the energy deposition spectra for a 2 GeV photon injected at $0^{\circ}$ to the crystal matrix. 
The dependence of energy release and energy resolution on the incoming photon energy is shown in Fig. \ref{Fig:2in1}. When a photon hits the center of middle crystal, the total deposited energy in the 5 $\times$ 5 crystal matrix is about $(97.72\pm 0.28)\%$. This value slightly declines as the photon incidence angle is increased, as shown in Fig. \ref{Fig:2in1}(a). In Fig. \ref{Fig:2in1}(b), at $ \Theta$ =$ 0^{\circ}$, the energy resolution varied from 2.37 \% at  50 MeV to 1.62 \% at 2 GeV. Moreover, the figure shows how the obtained energy resolution declines as the photon incidence angle is increased.\\
By using function $\sigma_{E}/{E}= a / \sqrt{E}\oplus b $, energy resolution parameters $a$ and $b$ at a range of $0^{\circ}$ to $15^{\circ}$ were calculated, see table  \ref{tab:table1}. As the photon incidence angle increased, a and b increased, which led to decreasing energy resolution. All these deteriorations  occurred  because of the shower leakages through the lateral and back surfaces of the crystal as shown in Fig. \ref{Fig:shower}, \ref{Fig:2in1}. Since the lateral leakages increased as the incidence angle rose, the results for energy resolution became incompatible with the fit function, resulting in some badly fitted curves. Energy resolution values obtained at $ 0^{\circ}$ are compatible with previous studies \cite{berra2014lyso}, \cite{saad2019impact}.

\begin{table}[h]
\caption{\label{tab:table1}Energy resolution fit parameters   for different photon incidence angles.}
\begin{tabular}{ccc} \hline \hline
Incident Angle  &{Stochastic Term a$(\%)$}&{Constant Term b$(\%)$}\\ \hline 
 $0^o$ & 0.42 $\pm$ 0.01  & 1.60  $\pm$ 0.02 \\ 
 $5^o$ & 0.47 $\pm$ 0.01  & 1.64  $\pm$ 0.03 \\
 $7^o$ & 0.48 $\pm$ 0.01  & 1.66  $\pm$ 0.02 \\
 $9^o$ & 0.49 $\pm$ 0.02  & 1.68  $\pm$ 0.03 \\
 $10^o$& 0.50 $\pm$ 0.02  & 1.71  $\pm$ 0.04 \\
 $15^o$& 0.51 $\pm$ 0.20  & 1.80  $\pm$ 0.04 \\ \hline \hline
\end{tabular}
\end{table}

\begin{figure}[htb]
\centerline{%
\includegraphics[width=12.5cm]{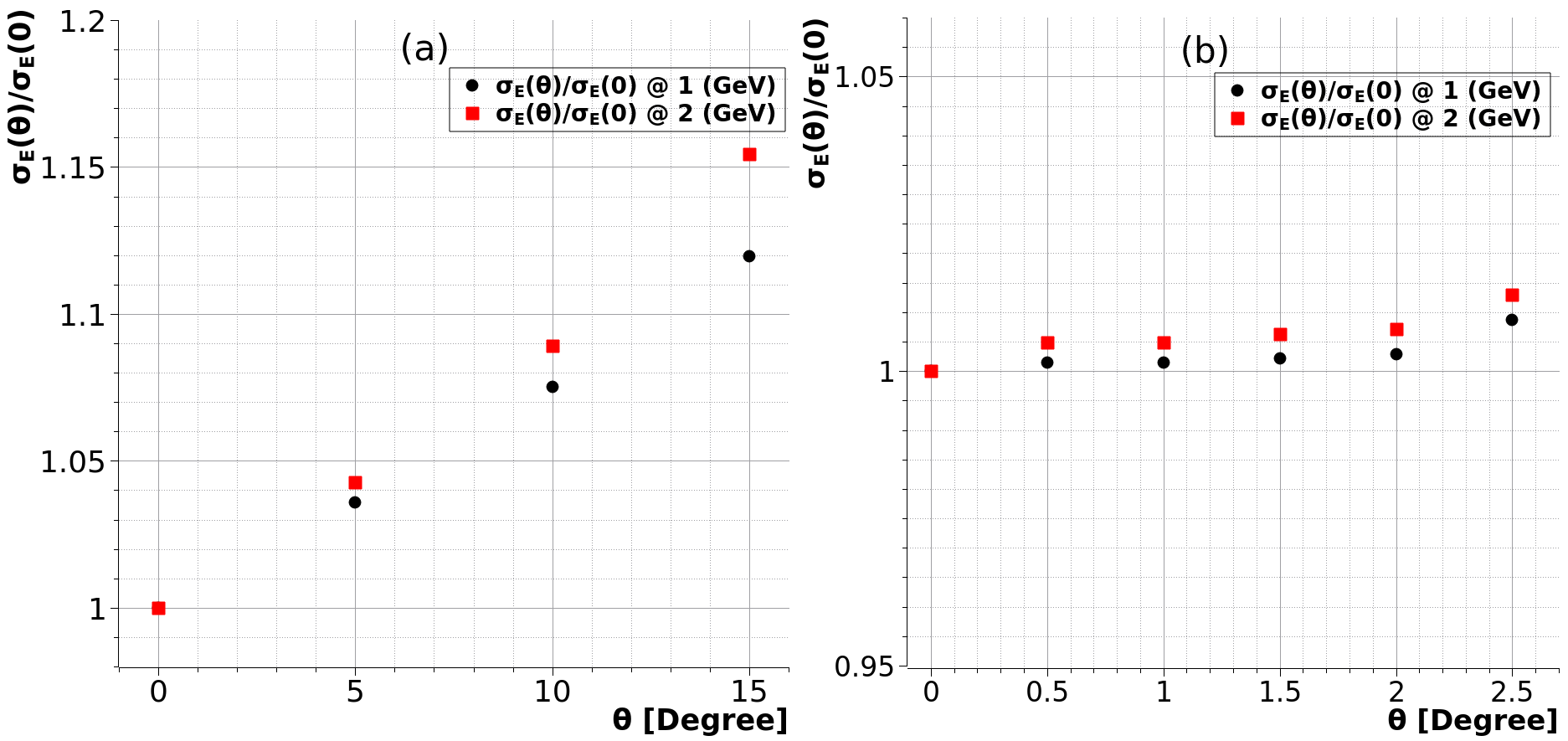}}
\caption{(a) Dependence of the simulated energy resolution on the angle of photon incidence. Resolution was normalized to  $0^o$. (b) Detailed view for $ 0^{\circ}$ to $ 2.5^{\circ}$.}
\label{Fig:2in1_2}
\end{figure}

\begin{figure}[htb]
\centerline{%
\includegraphics[width=10.5 cm]{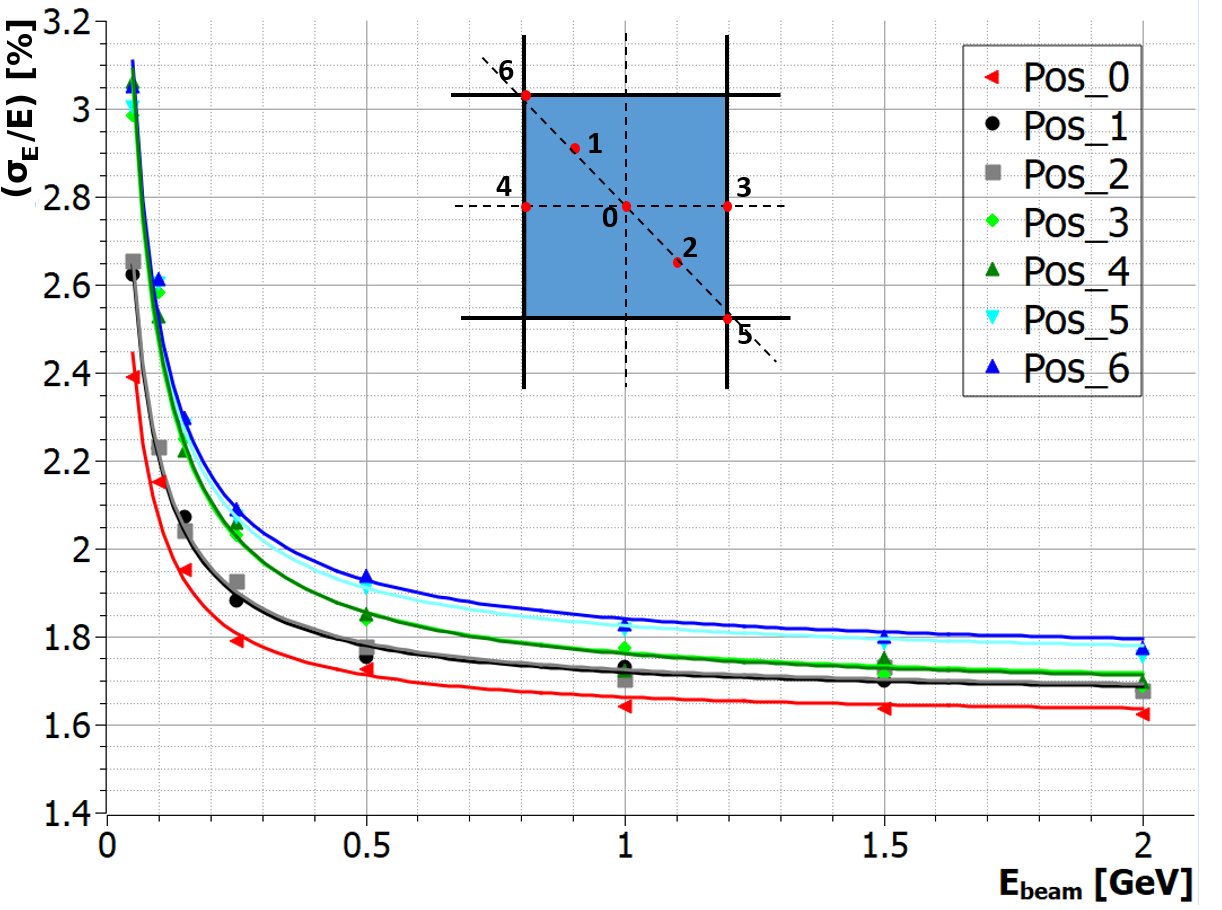}}
\caption{Energy resolutions  as a function of energy for different interaction positions at the central crystal of a 5 $\times$ 5 LYSO matrix.}
\label{Fig:sigma1}
\end{figure}

The simulation also shows that no changes were detected in energy resolution at small angles ($0^{\circ}$ to $2^{\circ}$), but it began to deteriorate at angles greater than $2^{\circ}$, see Fig. \ref{Fig:2in1_2}. Similar results were obtained in previous studies with the PWO crystals, which have similar radiation length and Moli$\grave{e}$re radius  as the LYSO crystal \cite{batarin2003precision}. To better understand the calorimetric energy resolution and to predict its behavior at various levels of energy and impact positions of the photon, different positions were scanned with the same module and Monte Carlo method. Fig. \ref{Fig:sigma1} shows the energy resolution as a function of the incident photon energy for the different interaction positions on the central crystal.

The energy resolution was highest value at the center of the central crystal (Pos\_0) and it began to deteriorate as the impact position of the photon moved away from the center of the crystal. Moreover, we obtained almost the same values for symmetrical positions, see Fig. \ref{Fig:sigma1}. Positions 1 and 2, 3 and 4, and 5 and 6 showed almost same energy resolution values. Position 1 and Position 2 showed better energy resolution than points on the edges and corners of the crystal, as there was be less lateral leakage because of their closeness to the center of the crystal matrix. 

Table \ref{tab:table2} shows energy resolution parameters (a and b) for different interaction positions of the incident photon. Results of the symmetric points indicated almost same a and b values. While contribution to the stochastic term of energy resolution from simulation was mainly due to transverse leakages, the contribution to the constant term was due to leakages from the back of the crystals and from transverse leakages. As seen in table \ref{tab:table2}, the contribution to energy resolution from the stochastic and constant term increased due to the transverse leakages as we moved away from the center of the crystal, so the resolution deteriorates.

\begin{table}[h]
\caption{\label{tab:table2}Energy resolution fit parameters for different interaction positions. The interaction positions 
0 to 6 are illustrated in Fig. \ref{Fig:crystal} and \ref{Fig:sigma1}.}
\begin{tabular}{ccc} \hline \hline
Interaction Position &{Stochastic Term a$(\%)$}&{Constant Term b$(\%)$}\\ \hline 
Pos\_0& 0.42 $\pm$ 0.01 & 1.60 $\pm$ 0.02  \\ 
Pos\_1& 0.46 $\pm$ 0.01 & 1.65 $\pm$ 0.01  \\
Pos\_2& 0.47 $\pm$ 0.01 & 1.65 $\pm$ 0.01 \\
Pos\_3& 0.58 $\pm$ 0.02 & 1.68 $\pm$ 0.03 \\
Pos\_4& 0.59 $\pm$ 0.01 & 1.67 $\pm$ 0.02 \\
Pos\_5& 0.57 $\pm$ 0.01 & 1.75 $\pm$ 0.03  \\
Pos\_6& 0.58 $\pm$ 0.01 & 1.76 $\pm$ 0.02 \\ \hline \hline
\end{tabular}
\end{table}

\newpage

\section{Conclusion}
In high-energy physics detectors, the precision of the electromagnetic calorimeters made of large crystals has immense importance due to their ability to identify photons and measure energy resolution for electrons and photons. In this paper, we studied the dependence of energy resolution on the photon incidence angle on a LYSO crystal calorimeter that was 25 $\times$ 25 mm$^{2}$ and 200 mm long. The calorimeter has been proposed for the ECAL of the TAC-PF detector. Calculations were made in the photon energy range from 50 MeV to 2 GeV. For 2 GeV photons, energy resolution was calculated as 1.62\% at $0^{\circ}$, and 1.87\% at $15^{\circ}$. Energy resolution did not change until the angle was about $2^{\circ}$. After that, it began to deteriorate. It could be stated that resolution will decrease significantly at an incidence angle of more than $2^{\circ}$ due to lateral and longitudinal  leakages from the crystal matrix. Furthermore, different interaction positions of the photon and their dependence on energy resolution were investigated. Energy resolution was worse at the corners of the central crystal than at the center of central crystal by a factor of 1.3 at 50 MeV and 1.1 at 2 GeV. The investigations presented in this paper are an important step towards  choosing the best crystal for ECAL of the TAC-PF detector. This study also can help researchers in the field of high-energy physics and detector design to optimize crystal reconstruction algorithms, which covers the effects associated with both angle of incidence and the interaction point of photons.  \\

The numerical calculations reported in this paper were partially performed at TUBITAK ULAKBIM, High Performance and Grid Computing Center (TRUBA resources).


\newpage
\bibliographystyle{unsrt}
\bibliography{./bibliography/template}

\end{document}